# Observation of a Deep Visual "Eclipse" in the WC9-Type Wolf-Rayet Star, WR 76

**Rod Stubbings**

*Tetoora Road Observatory, 2643 Warragul-Korumburra Road, Tetoora Road, 3821, Victoria, Australia; stubbo@sympac.com.au*

**Peredur Williams**

*Institute for Astronomy, Royal Observatory, Blackford Hill, Edinburgh, EH9 3HJ, United Kingdom; pmw@roe.ac.uk*

**Abstract**

The WC9-Type Wolf-Rayet star WR 76 is one of the most prolific dust makers identified from its infrared emission. WR 76 experienced a deep fading eclipse in 2016. The ~3.1 magnitude depth of the eclipse exceeds fadings in similar eclipses observed in WR stars thus far. Conclusions from recent and earlier analyses of eclipses observed suggests that WR 76 may be a prolific eclipser.

**1. Introduction**

Abrupt visual transient fading events, sometimes referred to as "eclipses"; a term that will be used throughout this paper, can be caused by the obscuration by dust in the line of sight. Such eclipses have been observed in R Coronae Borealis (R CrB) stars by Loreta (1934) and O'Keefe (1939) and classical novae (Gehrz 1988) for many decades, and more recently in WC-type Wolf-Rayet stars (Veen et al. 1998). All three sub-classes also frequently exhibit infrared (IR) emission from dust heated by the stellar radiation (Feast & Glass 1973; Gehrz 1988; Williams et al. 1987). These two manifestations of dust formation are complementary in the information they convey. The IR emission measures all the dust, not restricted to that in the line of sight, and can indicate the dust grain temperature and chemistry; which is usually carbon, but sometimes can be more complex. . This is evident in the following subclasses: R CrB stars (García-Hernández et al. 2013), novae (Helton et al. 2014) and WR stars (Chiar & Tielens 2001). On the other hand, the eclipses can give an indication of the sizes of grains if they have been observed in more than one passband (Veen et al.1998; Fahed et al. 2009),

From the IR light curves that show periodic outbursts, such as the prototype WR 140, ( Williams et al. 1990) and IR imaging that show rotating "pinwheel" structures in the prototype WR 104 ( Tuthill et al. 1999), dust formation is found to be related to wind interactions in WR+OB binary systems. The WR numbers for the Wolf-Rayet stars in this paper were assigned by van der Hucht et al. (1981) and van der Hucht (2001), who also give alternative designations.  In contrast, the incidence of eclipses by WR stars appears to be sporadic. The suggestion that the eclipses of WR 104 had the same period as the rotating dust pinwheel (Kato et al. 2002a) was not supported in a long-term monitoring study (Williams 2014) of dust-making WC stars based on the V-band photometry in the ASAS-3 survey (Pojmański 2002). The ASAS-3 data were accumulated from 2001 to 2009 which, allowing for seasonal and other gaps, provided data equivalent to around four years of continuous coverage for each star. The highest frequency of eclipse events was shown by WR 104, amounting 7/8 of the time, but the IR images of the dust pinwheel (Tuthill et al. 2008) indicate that eclipse-made dust was not an important contributor to its dust cloud. Several eclipses were also observed from WR 106 and other WC type stars in the ASAS-3 data.

WR 53 showed no optical eclipses in the 2001-2009 ASAS-3 data, and exhibited no variation or shallow eclipse activity. Independent visual and additional photometric monitoring of WR 53 commenced in 2011 to look for variations and surprisingly a deep 1-magnitude eclipse in 2015 was discovered (Stubbings 2015). WR 53 was also one of the WC8-9 stars known from its IR emission to be a dust maker (Williams et al. 1987), re-radiating about 3.6% of its UV-optical flux in the IR. The corresponding fraction for WR 104 was 60%, which led to the suggestion of a possible correlation between dust luminosity of the IR and the frequency of eclipses (Williams 2014).  Other stars known to be prolific dust makers from their IR emissions, WR 48a, WR76 and WR118 were too faint for the ASAS-3 survey to detect eclipses (Williams 2014). These stars were proposed for monitoring for visual eclipses to test the suggested correlation concerning dust luminosity of the IR and eclipses (Williams 2014). The brightest of these WR stars for visual monitoring is WR 76, (J2000: R.A. 16h 40m 05.25s, Dec. -45 41'12.7") and was measured at Vo = 15.46 (Fahed et al. 2009). WR 76 is a WC9 star (Torres et al. 1986) which re-radiates ~68% of its optical-UV luminosity in the IR (Williams et al. 1987), and this was selected for concentrated monitoring to search for eclipses in the present study.

**2 Observations**

WR 76 has a visual observing window from February to November. A visual monitoring program commenced in April 2016 to search for eclipses. A visual chart with comparison stars of 15.2, 15.5 and 16.1 were used for reference. After months of monitoring WR 76, which remained at around 15.2, a small dip was noticed on JD 2457606 (August 4) at a visual magnitude of 15.5. One of the authors (RS) contacted Peter Nelson to check WR 76 with a V-band image who was able to obtain an image and time series observations 3 days later giving V = 14.7 with no fading trend apparent. On the same night the author noticed WR 76 had returned back to around 15.3 visually. WR 76 remained at a visual magnitude of 15.3 until JD 2457653 (September 22) when again it declined to 15.5 and remarkably fainter to 16.0 the following night. A V- band image was obtained 4 days later on JD 2457658 (September 26) at V = 15.29. The data indicated a ~ 0.6 magnitude drop in V and ~ 0.7 visually, therefore a definite eclipse event was in progress. The fading trend continued and on JD 2457667 (October 6) a visual magnitude of 16.8 was recorded. At this point communication was made with Nidia Morrell (private communication 2016) to see if spectroscopy could be obtained from the El Leoncito Astronomical Complex astronomical observatory in Argentina. Unfortunately the elevation of WR 76 was then below the limit of their telescope. An additional V-band image was taken by Nelson on JD 2457773 (October 12) indicating WR 76 was now in a deep eclipse at V = 17.7.

With the full Moon and cloud cover making it difficult to observe visually, the next available observations were obtained on JD 2457685 (October 24) at a visual magnitude of 17.3 and also a V-band image at V =17.4. The rising branch of the eclipse progressed to maximum light on JD 2457702 (November 10) at a visual magnitude of 15.0. The eclipse commenced on JD 2457653 (September 22 ) and recovered on JD 2457702 ( November 10 ) with a duration of ~ 49 days and a depth of ~ 3.1 magnitudes in V and perhaps quite fainter due to the gap in data. The possibility of a slight minor eclipse on JD 2457606 (August 4) could also be linked the main eclipse event. The light curve is presented in Figure 1. The visual difference of 0.5 - 0.6 mag. at maximum to Johnson V could be due to spectral aspects of the variable's light and does not affect the result.

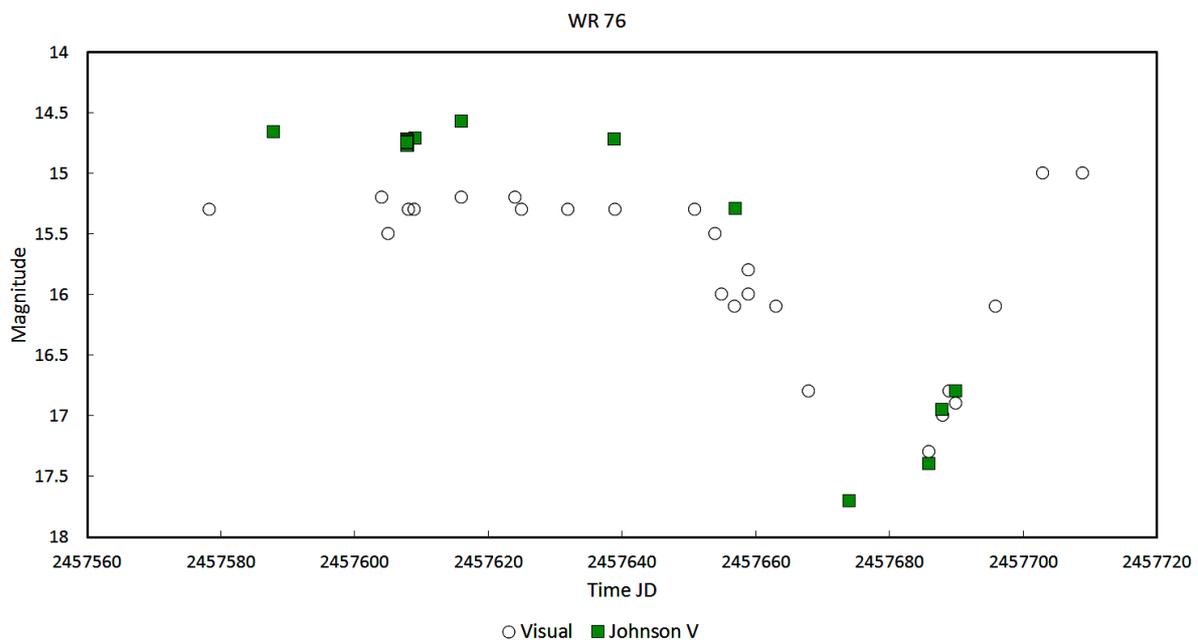

Figure 1. AAVSO light curve showing the deep ~3.1 magnitude eclipse beginning on JD 2457653 – 2457702 (September 22 – November 10) and lasting for 49 days before returning to maximum light. Black circles are visual data, green squares are Johnson V.

## 3. Discussion

The eclipse in WR76 observed in 2016 is deeper than those observed in WR 104 (Kato et al. 2002a; Williams 2014) and comparable to a 2.9 magnitude fading eclipse observed in WR 106 by Kato et al. (2002b) throughout an eight year monitoring period. Low level variation of $\sigma(v) = 0.06$ during a 1-month observing run had been observed in WR 76 by Fahed et al. (2009) and from comparison with variation in (v-i) was interpreted in terms of variable extinction by dust. Observations of WR 76 in the Bochum Galactic Disk Survey (Hackstein et al. 2015) showed a well-defined $\Delta r \sim 0.4$ magnitude eclipse in mid-2013 followed by another fading into a deeper eclipse at $\Delta r \sim 1.7$ magnitude and parts of other eclipses in shorter observing runs in 2012 and 2014 which are presented in Figure 2. Unfortunately, the cadence of the observations which were not dedicated to this star, but were surveying the whole field for variables, has not allowed for the deeper eclipse to be defined. The relative amplitudes $\Delta r/\Delta i \sim 1.2$ is also consistent with eclipsing by sub-micron particles (*cf.* Veen et al. 1998; Fahed et al. 2009), suggests that the visual eclipse reported in this paper has a similar origin.

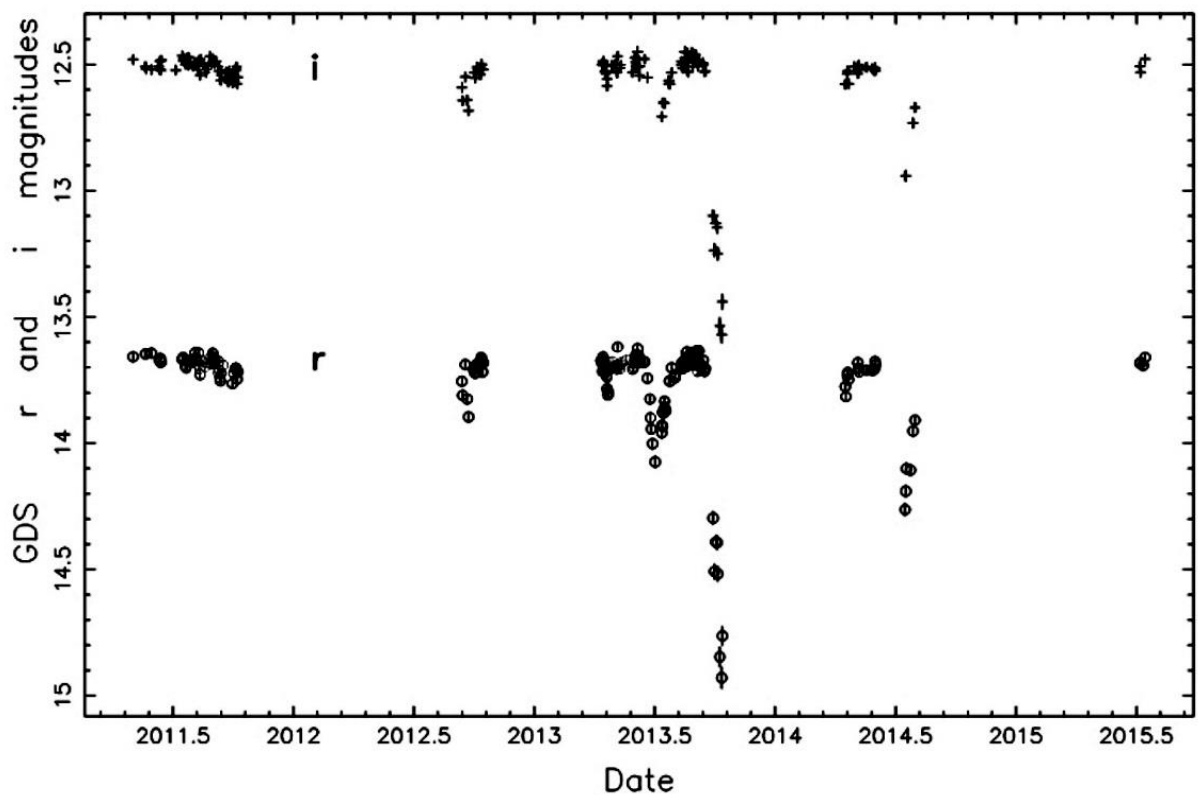

Figure 2. Observations of WR 76 in the Bochum Galactic Disk Survey showing recorded eclipses in r and i magnitudes.

In this case, the fading by about 3 mag. over approximately 25 days from JD 2457650 (September 18) would have been caused by the rapid formation of a cloud of dust particles in the line of sight, which covered the stellar disk. The dust that would have formed in the stellar wind of WR 76 is expected to have a terminal velocity near 1000 km/s (van der Hucht 2001), which would have dissipated the cloud and reduced its extinction. While WR 76 was fading the formation rate of the dust must have been high enough to more than compensate for its dissipation. When dust formation ceased and the existing dust was dissipated, the extinction fell and the flux recovered to its pre-eclipse level over a period of about 20 days. Veen et al. (1998) developed models to fit the eclipse light curves of three other WR stars: WR 103, WR 121 and WR 113, and found that the dust was forming significantly closer to the stars than the amorphous carbon dust clouds modelled from IR photometry by Williams et al. *(1987).* Support for the scale of the IR dust cloud models comes from detailed modelling of the dust pinwheel around WR 104 by Harries et al. *(2004).* This suggests that the dust grains modelled by Veen et al. (1998) would not have been able to survive heating in the stellar radiation field. On the other hand, independent evidence for the presence of scattering particles extremely close to the WC star component of WR 113 was provided by the observations of David-Uraz et al. (2012). This suggests that the particles responsible for the eclipses observed from WR stars may be more refractory and have optical properties very different to those of the amorphous carbon grains believed to be responsible for the IR emission. Further study of these phenomena would be helpful in understanding the formation of dust in the most hostile environments.

**4. Conclusion**

The combined results from recent and earlier studies of the deep eclipse observed only during one season of monitoring suggests that WR 76 may be a prolific eclipser, but further observations are required in order to test this and characterise these events.
In the four seasons in which WR 76 was observed in the Bochum Galactic Disk Survey, eclipses or parts of eclipses were observed in three out of the four, which also implies that eclipses may be rather frequent. The eclipse observed in WR 76 is one of the deepest known thus far compared to deep eclipses observed in WR 104 and WR 106. Eclipses found in WR stars are generally rare with only one detected eclipse from WR 53 in 14 years observing, apart from the heavier dust makers, suggesting WR 76 could be an excellent subject for the further study of this phenomenon.

**5. Acknowledgements**

We acknowledge with thanks the variable star observations from the AAVSO International Database (Kafka 2016) contributed by observers worldwide and used in this research. This research has made use of the International Variable Star Index (VSX) database (Watson *et al.* 2015), operated at AAVSO, Cambridge, Massachusetts, USA, and the Bochum Galactic Disk Survey. We would like to thank Peter Nelson for providing V-band images. We would also like to thank the anonymous referee whose comments and suggestions were helpful.

### References

Chiar, J. E., and Tielens, A. G. G. M., 2001, *Astrophys. J.,* **550**, L207.

David-Uraz, A., *et al.* 2012, *Mon. Not. Roy. Astron. Soc.*, **426**, 1720.

Fahed, R., Moffat, A. F. J., and Bonanos A. Z., 2009, *Mon. Not. Roy. Astron. Soc.,* **392**, 376.

Feast, M. W., and Glass, I. S., 1973, *Mon. Not. Roy. Astron. Soc.,* **161**, 293.

García-Hernández, D. A., Rao N. K., and Lambert, D. L., 2013, *Astrophys. J.*, **773**, 107.

Gehrz, R., D., 1988, *Ann. Rev. Astron. Astrophys.,* **26**, 377.

Hackstein, M., *et al.* 2015, *Astron. Nachr.*, **336**, 590 (data retrieved from table J/AN/336/590 at http://vizier.u-strasbg.fr).

Harries, T. J., Monnier, J. D., Symington, N. H., and Kurosawa, R., 2004, *Mon. Not. Roy. Astron. Soc.,* **350***,* 565.

Helton, L. A., Evans, A., Woodward, C. E., Gehrz, R. D., and Vacca, W., 2014, in *Stella Novae: Past and Future Decades*, ed. P. A. Woudt, V. A. R. M. Ribeiro, ASP Conf. Ser. 490, Astronomical Society of the Pacific, San Francisco, 261.

Kafka, S., 2016, variable star observations from the AAVSO International Database (https://www.aavso.org/aavso-international-database).

.Kato, T., Haseda, K., Yamaoka, H., and Takamizawa, K., 2002a, *Pub. Astron. Soc. Japan,* **54**, L51.

Kato, T., Haseda, K., Takamizawa, K., and Yamaoka, H., 2002b, *Astron. Astrophys.,* **393**, L69.

Loreta E., 1934, *Astron. Nachr.*, **254**, 151

O'Keefe J. A., 1939, *Astrophys. J.*, **90**, 294

Pojmański, G., 2002, *Acta Astron.,* **52**, 397.

Stubbings, R. 2015, *J. Amer. Assoc. Var. Star Obs.,* **43**, 163.

Torres, A. V., Conti, P. S., and Massey, P., 1986, *Astrophys. J.*, **300**, 379.

Tuthill, P. G., Monnier, J. D., and Danchi, W. C., 1999, *Nature,* **398**, 486.

Tuthill, P. G., Monnier, J. D., Lawrance, N., Danchi, W. C., Owocki, S. P., and Gayley, K. G., 2008, *Astrophys. J.,* 675, 698.

van der Hucht, K. A., 2001, *New Astron. Rev.*, **45**, 135.

van der Hucht, K. A., Conti, P. S., Lundstrom, I., and Stenholm, B., 1981, *Space Sci. Rev.,* **28**, 227.

Veen, P. M., van Genderen, A. M., van der Hucht, K. A., Li, A., Sterken C., Dominik, C., 1998, *Astron. Astrophys.,* **329**, 199.

Watson, C., Henden, A. A., and Price, C. A., 2015, *AAVSO International Variable Star Index VSX* (Watson+, 2006–2015; http://www.aavso.org/vsx).

Williams, P. M., van der Hucht, K. A., and Thé, P. S., 1987, *Astron. Astrophys,* **182**, 91.

Williams, P. M., van der Hucht, K. A., Pollock, A. M. T., Florkowski, D. R., van der Woerd, H., and Wamstecker, W. M., 1990, *Mon. Not. Roy. Astron. Soc.,* **243**, 662.

Williams, P. M., 2014, *Mon. Not. Roy. Astron. Soc.,* **445**, 1253.